\documentclass[a4paper]{jpconf}
\usepackage{graphicx}

\newcommand{\gev}{\,{\rm GeV}}

\begin{document}
\title{Probing GPDs in photoproduction processes at hadron colliders.}

\author{D Yu Ivanov$^1$, B Pire$^2$, L Szymanowski$^3$ and  J Wagner$^3$}

\address{$^1$ Sobolev Institute of Mathematics and Novosibirsk State University, 630090 Novosibirsk, Russia}

\address{$^2$ Centre de physique th\'eorique, \'Ecole polytechnique, CNRS, 91128 Palaiseau, France}

\address{$^3$ National Centre for Nuclear Research (NCBJ), Warsaw, Poland}

\ead{d-ivanov@math.nsc.ru}

\begin{abstract}

Generalized parton distributions (GPDs) enter QCD factorization theorems for hard exclusive reactions. They encode rich information about
hadron partonic structure. We explore a possibility to constrain GPDs in experiments at LHC considering two different exclusive processes:
the timelike Compton scattering  and  the photoproduction of heavy vector mesons.
\end{abstract}

\section{Introduction}

GPDs enter factorization theorems for hard exclusive reactions  in a similar manner as conventional parton distribution functions  (PDFs) enter factorization  for inclusive DIS\cite{gpdrev}. GPDs allow to access correlated information about the light cone momentum fraction and
transverse location of partons in hadrons measuring the transverse momentum dependence of the GPDs \cite{Burk}.
The studies of deeply virtual Compton scattering (DVCS)  $\gamma^*(Q^2)p\to \gamma p$ in electron-proton collisions
at JLAB, HERA are the primary source of our knowledge on GPDs. New data are expected from the planned experiments at  JLAB12, COMPASS,
and eventually at discussed electron-ion collider \cite{EIC}.

An interesting opportunity to access GPDs  is to consider hard exclusive processes at high energy hadron colliders, which are powerful
sources of quasi real photons in ultraperipheral collisions \cite{UPC}. The cross sections are  described with good accuracy
in the equivalent photon approximation (EPA) formula
\begin{eqnarray}
\sigma^{AB} =
\int dk_A
\frac{dn^A}{dk_A}\sigma^{\gamma B}(W_A(k_A)) +
 \int dk_B
\frac{dn^B}{dk_B}\sigma^{\gamma A}(W_B(k_B)) \nonumber
\end{eqnarray}
through: $\frac{dn}{dk}$-  an equivalent photon flux (the number of photons with energy $k$), and corresponding photoproduction cross sections.

The high luminosity and energies of these quasi-real photon beams, particularly at the LHC, open a new kinematical domain for the study of exclusive processes. It allows to determine sea-quark and gluon GPDs in the small skewedness region.
The golden channel to access GPDs in such quasi real photon processes is lepton pair production with a large invariant mass Q, either in the continuum or near a charmonium or bottomonium resonances such as $J/\Psi$ and $\Upsilon$. In the continuum, the process known as timelike Compton scattering (TCS) \cite{TCS,BGV}, $\gamma p\to \gamma^*(Q^2) p$, is the timelike analog of DVCS.  $J/\Psi$ and $\Upsilon$ production has the advantage of larger cross sections in corresponding regions of $Q$, but their theoretical description  includes  additional quantities related with  nonrelativistic quarkonium bound states.

In the case of TCS the all order proof of QCD collinear factorization at leading twist follows the same lines of arguments as the one for DVCS.
This solidly establishes the validity of the approach. The situation is more complicated in the case of heavy vector meson photoproduction.
Despite the lack of all order proof, it was demonstrated that
at leading twist (to leading
power in $1/m$ counting, where $m$- pole mass of the heavy quark) and up to  one--loop order in perturbation
theory the amplitude is given by the convolution of the perturbatively
calculable hard scattering amplitude and nonperturbative quantities.
The latter are
gluon and quark GPDs and the nonrelativistic QCD (NRQCD)
\cite{Bodwin:1994jh} matrix element $\langle O_1\rangle_V$ which
parametrizes an essential nonrelativistic dynamics of a heavy meson
system. This means that two firmly
founded QCD approaches, namely collinear factorization
and NRQCD, can be combined to construct a model free
description of heavy meson photoproduction.

\section{Timelike compton Scattering}

The calculable in perturbation theory coefficient functions of TCS and DVCS are related to one another by crossing.
The difference between coefficient functions in the timelike vs spacelike regimes can be traced back to the analytic structure (in the $q^2$ variable) of the scattering amplitude \cite{MPSW}. We found that $O(\alpha_s)$ corrections are rather large in timelike processes \cite{PSW,MPSSW}, and the results  are quite factorization scale dependent. Moreover, in the high energy region  strong dependence of predictions on the parametrization of gluon GPDs was demonstrated. It gives us a hope that this reaction could be used to constrain gluon and sea-quark  GPDs from the future experimental data.

As in the case of DVCS, a purely electromagnetic competing mechanism, the Bethe-Heitler (BH) mechanism contributes at the amplitude level to the same final state as TCS. BH and TCS contributions have different high energy asymptotics. With the growth of energy BH reaches a constant value, whereas
TCS contribution, related with QCD Pomeron, continues to grow. Nevertheless BH overdominates the TCS process in  most part of the  kinematical range available at LHC.  A winning strategy to obtain experimental information on TCS consists of choosing kinematics where the amplitudes of the two processes are of the same order of magnitude, and  using specific observables sensitive to the interference of the two amplitudes.

We estimated \cite{TCSUPC} the Born order lepton pair production in UPC at LHC, including both processes with various cuts to enable the study of the TCS contribution.
The pure Bethe - Heitler contribution to $\sigma_{p p}$, integrated over  $\theta = [\pi/4,3\pi/4]$, $\phi = [0,2\pi]$, $t =[-0.05 \gev^2,-0.25 \gev^2]$, ${Q}^2 =[4.5 \gev^2,5.5 \gev^2]$, and photon energies $k =[20,900]\gev $  gives:
\begin{equation}
\sigma_{pp}^{BH} = 2.9 \mbox{pb} \;.
\nonumber
\end{equation}
The leading order Compton contribution (calculated with NLO GRVGJR2008 PDFs, and $\mu_F^2 = 5 \gev^2$) gives:
\begin{equation}
\sigma_{pp}^{TCS} = 1.9 \mbox{pb}\;.
\nonumber
\end{equation}
We have choosen the range of photon energies in accordance with expected capabilities to tag photon energies
at the LHC. This amounts to a large rate of order of $\sim 10^5$ events/year at the LHC with nominal
luminosity ($10^{34}\,$cm$^{-2}$s$^{-1}$).

\begin{figure}
\centering
\includegraphics[keepaspectratio,width=0.45\textwidth,angle=0]{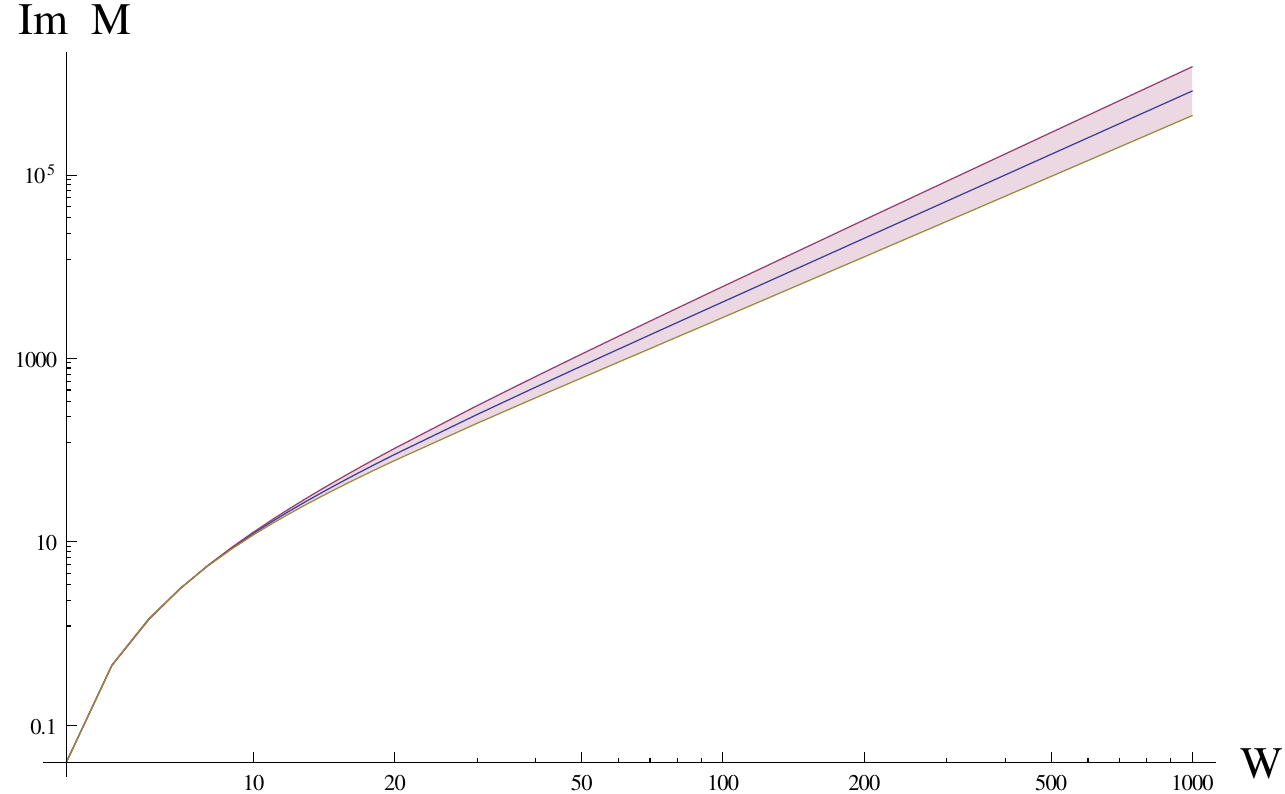}
\includegraphics[keepaspectratio,width=0.45\textwidth,angle=0]{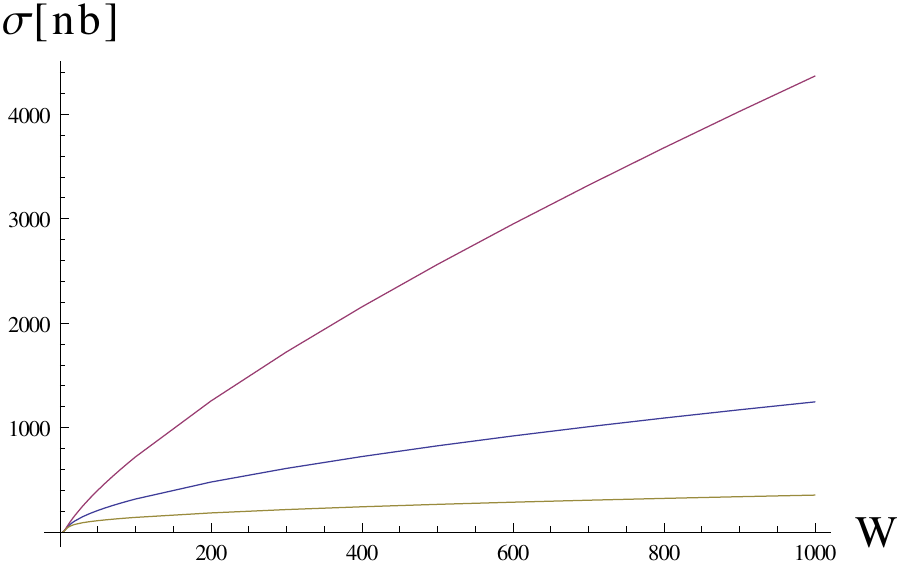}
\caption{(left) Imaginary part of the amplitude $\mathcal M$ and (right) cross section for photoproduction of $J/\psi$ at the LO as a function of $W = \sqrt{s_{\gamma p}}$ for $\mu_F^2 = M_{J/\psi}^2 \times \{0.5,1,2\}$ (respectively from bottom to top).
}
\label{fig:JPsi_LO}
\end{figure}
%

\section{Heavy Vector Meson Production}

The photoproduction of the heavy vector meson:
\begin{equation}
\gamma p \to V p
\label{VMprocess}
\end{equation}
is a subject of intense experimental \cite{VMexp} and theoretical \cite{VMth} studies. The main motivation of such studies is the possibility to explore gluon densities in the nucleon. In this work we present preliminary results \cite{Prep} on the use of the collinear factorization approach at the next to leading order in $\alpha_s$, which was developed in \cite{HVMP} , in the context of ultraperipheral collisons.

The amplitude of the process (\ref{VMprocess}) $\mathcal{M}$ is given by factorization formula:
\begin{eqnarray}
{\cal M}
&\sim &
\left(\frac{\langle O_1 \rangle_V}{m^3}\right)^{1/2}
 \int\limits^1_{-1} dx
\left[\, T_g( x,\xi)\, F^g(x,\xi,t)+
T_q (x,\xi) F^{q,S} (x,\xi,t) \,
\right] \, ,
\\
F^{q,S} (x,\xi,t)&=&\sum_{q=u,d,s}  F^q (x,\xi,t) \, ,
\end{eqnarray}
where   $\langle O_1 \rangle_V$ is given by the NRQCD through leptonic meson decay rate,
\begin{equation}
\Gamma[V\to l^+l^-]=\frac{2e_q^2\pi\alpha^2}{3}
\frac{\langle O_1\rangle_V }{m^2}
\left( 1-\frac{8\alpha_s}{3\pi}\right)^2 \, ,
\label{decay}
\end{equation}
$\xi$ is the fraction of the longitudinal momentum transfer, $t$ is the momentum transfer squared. $F^{q,g}$ are quark and gluon GPDs and $T^{q,g}$ are hard scattering coefficient functions given by:


\begin{equation}
 T_g(x,\xi)=\frac{\xi}{(x-\xi+i\varepsilon)(x+\xi-i\varepsilon)}
{\cal A}_g\left(\frac{x-\xi+i\varepsilon}{2\xi}\right) \quad , \quad
T_q( x,\xi)={\cal A}_q\left(\frac{x-\xi+i\varepsilon}{2\xi}\right) \, .
\label{gAT}
\end{equation}
At the leading order they read:
\begin{equation}
{\cal A}_g^{(0)}(y)=\alpha_s \quad , \quad
{\cal A}_q^{(0)}(y)=0 \, .
\label{LOq}
\end{equation}

In Fig.\ref{fig:JPsi_LO} we present the leading order result for $J/\psi$ photoproduction using the Goloskokov-Kroll model of GPDs \cite{GK}.
On the left panel of this figure we show the imaginary part of the amplitude $\mathcal{M}$, plotted for the following values of factorization scale $\mu_F^2 = M_{J\psi}^2\cdot\{0.5,1,2\}$ (respectively from bottom to top). The factorization scale dependence of the LO result is visible even more dramatically on the right panel of Fig.\ref{fig:JPsi_LO}, where the cross section is plotted. For the  high values of the center of mass energy $W$ the results for $\mu_F^2 = 0.5  M_{J\psi}^2$ and $\mu_F^2 = 2  M_{J\psi}^2$ differ by one order of magnitude.

%

\begin{figure}
\centering
\includegraphics[keepaspectratio,width=0.65\textwidth,angle=0]{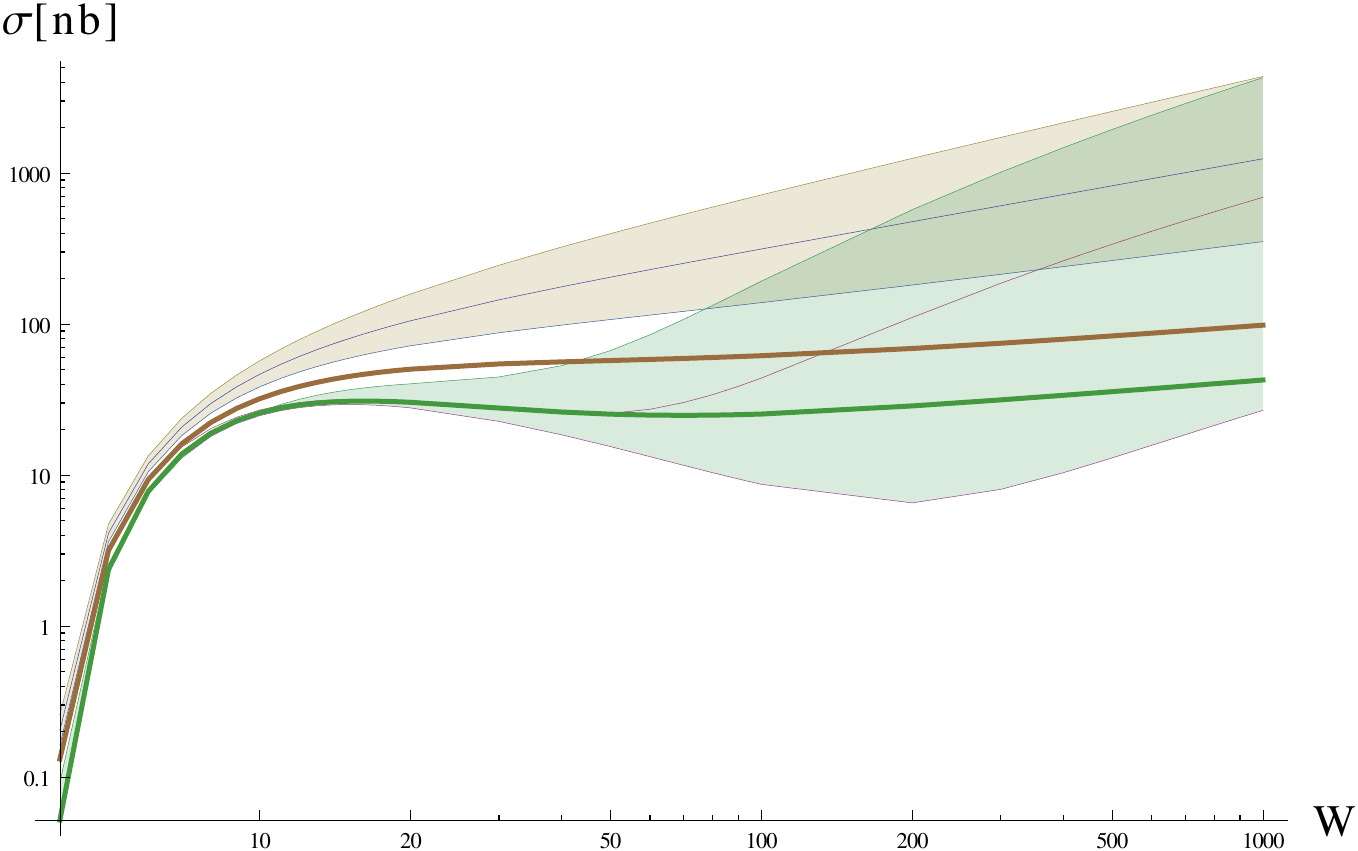}
\caption{$J/\psi$ photoproduction cross section as a function of $W = \sqrt{s_{\gamma p}}$ for $\mu_F^2 = M_{J/\psi}^2 \times \{0.5,1,2\}$- LO (grey band) and NLO (green band). Thick lines for LO(brown) and NLO(green) for $\mu_F^2 = 1/4  M_{J/\psi}^2 $. }
\label{fig:JPsi_NLO}
\end{figure}
%
%
%

Results for the NLO cross section are presented on the Fig.\ref{fig:JPsi_NLO}. Grey(green) band presents the results for LO(NLO) cross section for the factorization scale in the range $\mu_F^2 = M_{J\psi}^2\cdot\{0.5,1,2\}$. We see that the NLO corrections are very big and the overall result depends very strongly on the choice of the factorization scale, especially  for the high values of $W$.\footnote{In our calculations, in both LO and NLO cases, we keep the value of renormalization scale fixed $\mu_R=M_{J\psi}$.}

Why NLO corrections are so large in this case,  $\xi\ll 1$?  The inspection of
NLO hard-scattering amplitudes shows that the imaginary part of the amplitude dominates and
that the leading contribution to the NLO correction originates from the broad integration region
 $\xi\ll x\ll 1$ , where the gluonic part approximates ($\bar \alpha_s=3\alpha_s/\pi$):
\begin{equation}
Im {\cal M}^g\sim H^g(\xi,\xi)+\bar \alpha_s \left[\log\frac{M_V^2}{\mu_F^2}-\log 4\right]
\int\limits^1_{\xi}\frac{dx}{x} H^g(x,\xi)\, .
\label{C1}
\end{equation}
Given the behavior of the gluon GPD at small $x$,
$H^g(x,\xi)\sim xg(x)\sim \rm{const}$, we see that NLO correction is parametrically large, $\sim \ln(1/\xi)$, and negative unless one chooses the value of the factorization scale sufficiently lower than the hard kinematic scale, $Q=M_V$. Similar observations were reported also in the recent paper \cite{JMRT}.

The size of the corrections, and the sensitivity of the NLO result to the factorization scale choice, shows that some additional information about still higher order contributions  is needed to provide reliable theoretical predictions.
This may come in terms of some scale fixing procedure, like BLM method \cite{BLM} (assuming $\mu_R=\mu_F$), or some idea of the scale choice to minimalize the one-loop corrections (this possibility is shown on the Fig.\ref{fig:JPsi_NLO} - thick brown (green) line corresponds to the LO (NLO) result for $\mu_F^2 = 1/4 M_V^2$), the second approach was adopted in \cite{JMRT}.

In our opinion the most promising approach is related with the resummation of the higher orders terms enhanced at small $\xi$ by the powers of large
logarithms of energy, $\sim \bar \alpha_s^n \ln^n (1/\xi)$, see
\cite{Dima:Blois}:
\begin{equation}
{\cal I}m {\cal M}^g\sim H^g(\xi,\xi)
+ \int\limits^1_{\xi}\frac{d x}{x} H^g(x,\xi)
\sum\limits_{n=1}C_n(L)\frac{\bar \alpha_s^n}{(n-1)!}\log^{n-1}\frac{x}{\xi}\, ,
\end{equation}
here
$C_n(L)$ - polynomials of $L=\ln(M_V^2/\mu_F^2)$ which maximum power is $L^n$.
For DIS inclusive structure functions $F_T$ and $F_L$ corresponding $C_n(L)$ coefficients were calculated long time ago
by Catani and Hautman \cite{Catani:1994sq}. Their method developed for  inclusive DIS can be
straightforwardly generalized to exclusive, nonforward processes.

Resummed coefficient functions, parameterized by  $C_n(L)$ polynomials,  could be calculated and conveniently represented
using Mellin transformation and Mellin variable $N$. In the Mellin space the resummed coefficient function
is a polynomial in variable $z=\bar \alpha_s/N$. At the reverse transform the contributions proportional to the powers of this variabe, $\sim z^n$,
would generate terms $\sim \bar \alpha_s^n \ln^n (1/\xi)$ in the process amplitude.
For meson photoproduction our result in the  $\overline{MS}$ scheme reads
$$
1+
z (L-\ln 4)+
\frac{z^2}{6}\left(\pi^2+3\ln^2 4+3L(L-\ln 16)\right)+
 \dots \, ,
$$
where only two nontrivial terms of the expansion are shown (the methods allows to calculate any fixed number of such high energy terms).
So that, $C_1(L)=L-\ln 4$, in accordance with the found high energy asymptotic of NLO result in Eq. (\ref{C1}),
$C_2(L)=\left(\pi^2+3\ln^2 4+3L(L-\ln 16)\right)/6$,
and so on.

Let us discuss  the value of resummed coefficient and its  dependence  on the scale of factorization $\mu_F$. Below we present results for two cases.
a) $\mu_F$ is equal to kinematic hard scale $\mu_F=M_V$ ($L=0$), and b)  $\mu_F$ is chosen to vanish the value of the first high energy term $C_1$, it requires $\mu_F=M_V/2=m$ ($L=\ln 4$):
\begin{eqnarray}
&
a) \ \ \ (\mu_F=M_V): \quad   1 -1.39\, z +2.61\, z^2 +0.481\, z^3 -4.96\, z^4 +\dots
& \nonumber \\
&
b) \ \ \ (\mu_F=M_V/2): \quad  1 +0.\, z +1.64\, z^2 +3.21\, z^3 +1.08\, z^4 +\dots \, .
& \nonumber
\end{eqnarray}
We see that almost all the high energy term coefficients, $C_n(L)$, have large absolute values. It shows that it is important
to take into account not only large NLO effects but also contributions from still higher orders of QCD collinear expansion that are enhanced
by the powers of large logarithms of energy. Another important observation is that it is not possible by appropriate choice of
factorization scale $\mu_F$ to move all enhanced by powers of $\ln(1/\xi)$ contributions from the coefficient function into the GPD (through
its $\mu_F$- evolution). Such a strategy is promoted in \cite{JMRT}. We see that our results above, the  case b), do not support this suggestion.
The choice $\mu_F=M_V/2$ indeed eliminates the big part of NLO correction from the hard coefficient, but it can not allow to get rid of such big terms
from the higher orders contributions.

We believe that high energy resummation described above have to be incorporated in the analysis of
the  $J/\Psi$ and $\Upsilon$ meson photoproduction processes, this work is now in progress.

\section{Summary}
GDPs enter factorization theorems for hard exclusive reactions  in a similar manner as PDFs enter factorization for inclusive DIS.
To this time DVCS remains the main source of information about GPDs. A lot of new DVCS experiments are planned at JLAB12, COMPASS, eventually at EIC.
Ultraperipheral collisions at hadron colliders open a new way to measure GPDs in  TCS and photoproduction of heavy vector mesons.
NLO corrections were studied in the kinematics typical for experiments at LHC collider. We found rather large for TCS and even more
dramatic NLO effects for the case of vector meson  photoproduction.
High energy resummation was suggested as a tool to provide reliable theoretical predictions in the region of high energy.

\section{Acknowledgments}
This work is partly supported by the Polish Grant NCN No DEC-2011/01/D/ST2/02069, by the COPIN-IN2P3 Agreement and by the grant RFBR-15-02-05868.

\end{document}